\documentclass[11pt,twoside]{article}

\newcommand{\ds}{\displaystyle}

\newcommand{\beq}{\begin{equation}} 
\newcommand{\eeq}{\end{equation}} 
\usepackage{asp2010}
\usepackage{graphicx}
\usepackage{color}

\resetcounters 

\begin{document}

\resetcounters

\title{Towards a self-consistent orbital evolution for EMRIs}

\author{Alessandro Spallicci$^1$, Patxi Ritter$^{1,2}$, Sylvain Jubertie$^3$, St\'ephane Cordier$^2$, and Sofiane Aoudia$^4$}
\affil{$ö1$ 
Universit\'e d'Orl\'eans, Observatoire des Sciences de l'Univers OSUC,\\ 
Laboratoire de Physique et Chimie de l'Environnement et de l'Espace LPC2E, UMR CNRS 6115, 
3A Av. Recherche Scientifique, 45071 Orl\'eans, France}
\affil{$ö2$
Universit\'e d'Orl\'eans, 
Math\'ematiques - Analyse, Probabilit\'es, Mod\`elisation - Orl\'eans MAPMO, UMR CNRS 7349, 
Rue de Chartres, 45067 Orl\'eans, France}
\affil{$ö3$
Universit\'e d'Orl\'eans, Laboratoire d'Informatique Fondamentale d'Orl\'eans LIFO, EA 4022, 
Rue Leonardo da Vinci, 45067 Orl\'eans, France}
\affil{$ö4$ 
Max Planck Institut f\"ur Gravitationphysik, A. Einstein, \\
Am M\"uhlenberg 1, 14476 Golm, Deutschland}

\begin{abstract} 
We intend to develop part of the theoretical tools needed for the detection of gravitational waves coming from the capture of a compact object, 1-100 $M_\odot$, by a Supermassive Black Hole, up to a $10^9~M_\odot$, located at the centre of most galaxies. The analysis of the accretion activity unveils the star population around the galactic nuclei, and tests the physics of black holes and general relativity. 
The captured small mass is considered a probe of the gravitational field of the massive body, allowing a precise measurement of the particle motion up to the final absorption. The knowledge of the gravitational signal, strongly affected by the self-force - the orbital displacement due to the captured mass and the emitted radiation - is imperative for a successful detection.
The results include a strategy for wave equations with a singular source term for all type of orbits. We are now tackling the evolution problem, first for radial fall in Regge-Wheeler gauge, and later for generic orbits in the harmonic or de Donder gauge for Schwarzschild-Droste black holes.  
In the Extreme Mass Ratio Inspiral, the determination of the orbital evolution demands that the motion of the small mass be continuously corrected by the self-force, i.e. the self-consistent evolution. At each of the  integration steps, the self-force must be computed over an adequate number of modes; further, a differential-integral system of general relativistic equations is to be solved and the outputs regularised for suppressing divergences. 
Finally, for the provision of the computational power, parallelisation is under examination. 
\end{abstract}

\section{How motion of a particle is affected by its own mass and the emitted radiation}

A particle, of $z^\alpha$ coordinates, follows the geodesic given by

\beq
\frac{Du^\alpha}{d\tau}= \frac{du^\alpha}{d\tau} + ^{\rm b}\!\Gamma^\alpha_{\mu\nu} u^\mu u^\nu = 0 ~~,                                                
\label{bg}
\eeq
where $\tau$, $^{\rm b}\Gamma^\alpha_{\mu\nu}$, $u^\alpha\!\equiv\! dz^\alpha/d\tau$ are the proper time, Christoffel symbol and four-velocity in the background (b) metric $g_{\mu\nu}$, respectively. Let us now consider the same particle moving in a perturbed metric. 

In the restricted two-body problem,  \cite{springer2011}, the particle infinitesimal size implies that the perturbations diverge at the particle itself.
\cite{dw2003} adapted Dirac's approach to the self-force equation - the MiSaTaQuWa equation from \cite{mino1997}; \cite{quinn1997}. In flat spacetime, the radiative Green function is obtained by subtracting the singular contribution, half-advanced plus half-retarded, from the retarded Green function. The singular part does not exert any force on the particle, upon which only the regular field acts.
In curved spacetime, the attainment of the radiative Green function passes through the inclusion of an additional, purposely built, function $H$.  This approach emphasises that the motion is a geodesic of the full (f) metric ${\hat g}_{\mu\nu} = g_{\mu\nu} + h_{\mu\nu}^R$ where 
$h_{\mu\nu}^R$ is the radiative part of the perturbations, and it implies two notable features: the regularity of the radiative field and the
avoidance of any non-causal behaviour. The radiative $R$ component is conceptually given by

\beq
R = {\rm Ret} - {\rm Sing} = {\rm Ret} - \frac{1}{2}[{\rm Ret} +  {\rm Adv} - H] = \frac{1}{2}[{\rm Ret} - {\rm Adv} + H]~~,
\eeq
where the {\it ad hoc} function $H$ is defined to agree with the
advanced Green function when the particle is in the future of the evaluation point ($H$ = Adv); and to the retarded 
Green function when the particle is in the past of the evaluation point ($H$ = Ret), but differs from zero in the intermediate values of the world-line outside the light-cone. Thus, the radiative component includes the state of motion at all times prior to the advanced time and it is not a representation of the physical field, but rather of an effective field. Indeed, $H$ goes to zero when the evaluation point coincides with the particle position.   

We define 
$\hat{z}^\alpha = z^\alpha + \Delta z^\alpha$ as the coordinates of the particle in the full metric. The geodesic is given by

\beq
\frac{D\hat{u}^\alpha}{d\lambda}= 
\frac{d\hat{u}^\alpha}{d\lambda} + ^{\rm f}\!\Gamma^\alpha_{\mu\nu} \hat{u}^\mu \hat{u}^\nu = 0~~,
\label{fg}
\eeq
where $\lambda$, $^{\rm f}\Gamma^\alpha_{\mu\nu}$, $\hat{u}^\alpha\!\equiv\! d\hat{z}^\alpha/d\lambda$ are the proper time, Christoffel symbol and four-velocity in the full metric, respectively. We wish to compute the difference between the two geodesics, knowing that the final equation of motion of the particle in the perturbed background is given by $
a_{total} = {D^2 z^\alpha}/{d\tau^2}+ 
{D^2 \Delta z^\alpha}/{d\tau^2}$. 
Obviously, the gauge freedom allows to choose a comoving coordinate frame where no acceleration occurs. After some considerable manipulation, we get

\beq
\frac{D^2 \Delta z^\alpha}{d\tau^2}=
\underbrace{- {R_{\mu\beta\nu}}^\alpha u^\mu \Delta z^\beta u^\nu}_{Background~geodesic~deviation} 
\underbrace{- \frac{1}{2}
(g^{\alpha\beta} + u^\alpha u^\beta) 
(2h_{\mu\beta ;\nu}^R- h_{\mu\nu ;\beta}^R) u^\mu u^\nu}_{Self-acceleration~~MiSaTaQuWa}~~. 
\label{gweq}
\eeq
Stemmed from geodesic principles, an exact geodesic deviation equation at first order is obtained by subtracting the background from the perturbed motion, equation (\ref{gweq}). The first right-hand side term depends on the background metric, while the second depends upon the perturbations generated by the particle mass $m$, and it is the non-trivial MiSaTaQuWa self-acceleration. 
\cite{gralla2008} adduce that a first order perturbation scheme will let grow away from the exact solution at late times, and that no different destiny will occur to a second or higher order scheme at even later times. They assert that it is preferable i) to drop searching higher order self-force expressions; ii) to evolve the trajectory by continuously and iteratively applying the correction given by the second term, while disregarding the first term.  

The self-force is defined in the harmonic or de Donder (dD) gauge, where the ten metric components aren't combined into a wave equation, as in the Regge-Wheeler (RW) gauge. But, computation in other gauges, \cite{barack2001}, it is often not possible, as the variation due to the change from dD to a new gauge (G), $\delta F_{self}^{(dD\rightarrow G)}$ does not admit a well defined value. One exception is constituted by the radial trajectory, where the two self-forces (dD and RW gauges) can be made equal.    
The regularisation process subtracts the diverging or singular part (represented by the regularisation parameters $A^{\alpha }_{\pm}, B^{\alpha }, C^{\alpha}, D^{\alpha }$, which are gauge independent, and to be computed in the dD gauge) from the full perturbations, following 

\beq
F^{\alpha{\rm {(G)}}}_{\rm self}=\sum_{\ell=0}^{\infty }\left(F^{\alpha
\ell{\rm (G)}}_{\pm\rm full}-A^{\alpha }_{\pm}L-B^{\alpha }-C^{\alpha}/L\right)
-D^{\alpha }~~,  
\label{eqVI170}
\eeq
where $L = \ell + 1$, $\ell$ indicating the mode, and $\pm$ represents the two sides at the particle coordinate. For the non-adiabatic radial fall (radial coordinate $r$ and particle position in the background, $r_p$), in RW gauge and in coordinate time, the expression corresponding to the self-force is given by \cite{spallicci2004}

\beq
\Lambda_2=\sum_{\ell=0}^\infty\left(\Lambda_{\pm 2}^\ell-\tilde{A}^\alpha_{\pm} L-\tilde{B}^\alpha-\tilde{C}^\alpha/L\right)-\tilde{D}^\alpha~~,
\eeq
where $\tilde{A}^{\alpha }_{\pm}, \tilde{B}^{\alpha }, {\tilde C}^{\alpha}, {\tilde D}^{\alpha }$ are derived from equation \ref{eqVI170} and the corresponding untilded regularisation parameters, and  

{\small
\[\Lambda^\ell_{\pm 2} = \ds\sqrt{\frac{2\ell+1}{4 \pi}}\left\{\ds\frac{1}{r - 2M}
\left[
\frac {r^2 H^\ell_{\pm 2,t}}{2(r - 2M)} - \ds\frac {MH^\ell_1}{r - 2M} - rH^\ell_{\pm 1,r} 
\right]
\dot{r}_p^3
- \ds\frac {3}{2} H^\ell_{\pm 2,r}
\dot{r}_p^2
- 3 \left(
{ \ds\frac {H^\ell_{\pm 2,t}}{2}}  - { \ds\frac {MH^\ell_1}{r^2}}
\right )
\dot{r}_p\right.
\]
\beq
\left.
+ { \ds\frac {r - 2M} {r}} 
\left [ {\ds\frac{2MH^\ell_2}{r^2}} + 
{\ds\frac{(r-2M)H^\ell_{\pm 2,r}}{2r}} - H^\ell_{\pm 1,t} 
\right]\right\}~~,
\eeq}
being $M$ the black hole mass, $\dot{r}_p$ the particle velocity, $H^\ell_{1,2}$ perturbations (of $C^0$ continuity class) drawn by the gauge-invariant Moncrief wave function $\psi$.
The latter is derived from the Regge-Wheeler-Zerilli wave equation ($V^\ell$ potential, $r^*$ tortoise coordinate)

\beq \ds
{\left[-\frac{\partial^2}{\partial t^2}+\frac{\partial^2}{\partial r^{*2}}-V^\ell(r)\right ]\psi^\ell(r,t)=F^\ell(r)\delta\left (r-r_p(t)\right)+ 
G^\ell(r)\frac{\partial}{\partial_r}\delta\left (r-r_p(t)\right )}
~~.
\eeq

\begin{figure}
 \begin{minipage}[b]{.5\linewidth}
  \centering\includegraphics[width=0.7\linewidth]{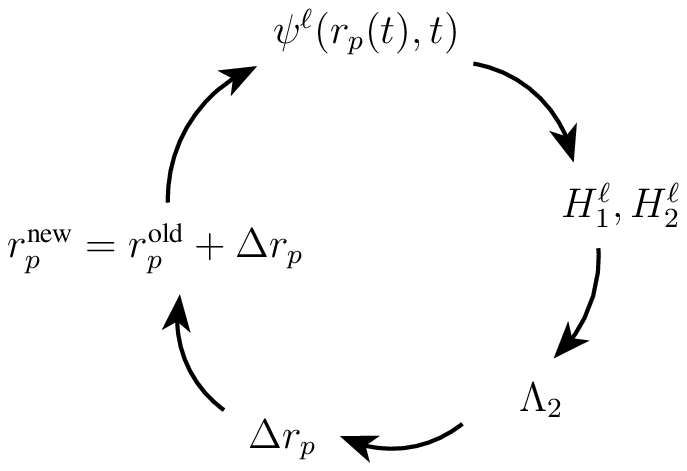}
  \caption{Iteration scheme for computation of the evolving orbit.}
	\label{fig.iter}
 \end{minipage} \hfill
 \begin{minipage}[b]{.5\linewidth}
  \centering\includegraphics[width=0.7\linewidth]{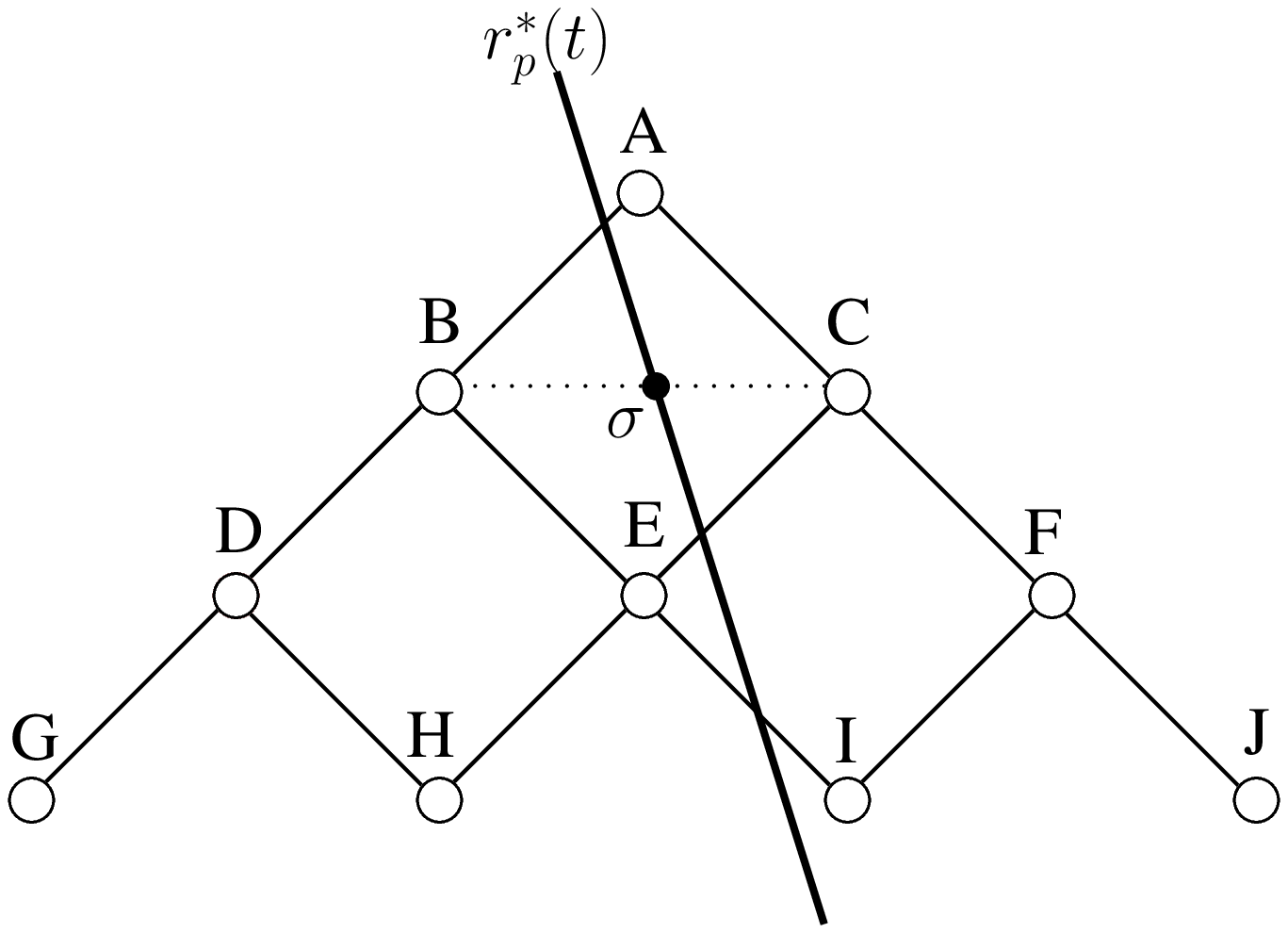}
  \caption{Fourth order scheme.}
	\label{4thorder}
 \end{minipage}
\end{figure}

The perturbations, and thereby $\Lambda_2$, depend upon $m$ the mass of the particle-star.    
The back-action shows as a correction ${\Delta r_p}$, that is ${{\hat r}_p(t)=r_p(t)+\Delta r_p(t)}$, and it obeys to a $t$-ODE, corresponding to equation \ref{gweq}

\beq 
\Delta \ddot{r}_p=\Lambda_0 (g_{\mu\nu}, r_p, {\dot r}_p)\Delta\dot{r}_p + 
\Lambda_1 (g_{\mu\nu}, r_p, {\dot r}_p) \Delta r_p+\Lambda_2 (h_{\mu\nu}, r_p, {\dot r}_p)~~.
\label{gweq-corr}
\eeq

The iterative approach, figure \ref{fig.iter}, demands an accurate reinterpretation of equation \ref{gweq-corr}.  
Firstly, for an infinitesimal time step, $\Lambda_0$ and $\Lambda_1$ vanish. Secondly, the ${\Lambda_2}$ parameter is to be computed on the new trajectory: indeed, $\Delta r_p$ represents here the difference with the trajectory computed at the previous integration step, and not anymore the background trajectory at start. Thirdly, each single iterated position and velocity may be identified with the coordinates of a particle possessing the same values and moving on a - to be determined - geodesic. This approach sums up the effects computed on successive osculating orbits, i.e. stretches of geodesics. 
  
\section{The algorithm}

Classical finite difference methods have to be adapted to deal with the discontinuity of the wave function ${\psi}$ and its derivatives on the trajectory ${r_p(t)}$ due to the infinitesimal size of the particle. Analytically derived jump conditions on ${\psi}$ and derivatives are used as guideline and reference throughout the integration, \cite{aoudia2011}; \cite{ritter2011}.
Fourth order accuracy on $\psi$ has been reached to compute the metric perturbations and their first derivatives (thereby implying third order derivatives of ${\psi}$)

\beq
{\displaystyle\psi^\ell_A=\sum_i\left[q_i\psi^\ell_i+\sum_{n+m<4}{\tilde{q}_iT^{(n,m)}_{i}[\partial^n_{r*}\partial^m_t\psi^\ell]_\sigma}\right]}+\mathcal{O}(h^5)~~,
\eeq

\beq
{\displaystyle [Q^{nm\ell}]_\sigma=\lim_{r\to r_p^+(t_\sigma)}Q^{nm\ell}-\lim_{r\to r_p^-(t_\sigma)}Q^{nm\ell}}~~,
\eeq

\beq
{\rm if} r^*(t_i)<r_p^*(t_i):\ \tilde{q}_i=0\ ,\ {\rm else}\ \tilde{q}_i=q_i~~,
\eeq
for ${i=\{B,C\dots J\}}$, ${ Q^{nm\ell}(r,t)=\partial^n_{r*}\partial^m_t\psi^\ell(r,t)}$, $T^{(n,m)}_{i}$ are Taylor coefficients and $q_i$ are constants depending on the way the particle crosses the cells, Figure \ref{4thorder}.

\section{Parallel computing}

Parallelisation allows better performance, in terms of resolution and processing time, and it is an evident aid for the computation of 
orbital evolution. At this preliminary stage though, only the non-iterative code has been worked upon.  
The availability of parallel hardware doesn't imply an immediate exploitation of its capacity, as a simulation program often needs   refurbishment. The original sequential algorithm was improved by using loop unrolling and cache optimisation. The modified version runs  seven times faster, and it is used as standard reference.
The following parallel techniques have been investigated and tested on a machine equipped with two quad-core AMD Opteron running at 2.3GHz.

{\bf SSE instructions}. 
The SSE (Streaming SIMD - Single Instruction, Multiple Data - Extension) technology works with double-precision floating-point instructions applied onto a single arithmetical operation simultaneously, thus doubling the computational efficiency.
However, it requires to explicitly deal with the operations between the main memory and the processor SSE registers, while taking care of the memory alignment constraints for efficiency. This implies the redesign and rewriting of the algorithms for those instructions.
On one core, the SSE implementation achieves a 1.6 speedup over the reference implementation.
A speedup of 2 wasn't achieved, since the bus between the main memory and the processor was left unaltered, and it was unable to feed the SSE registers quickly enough to reach peak performance.

{\bf SSE instructions + Multi-Threading}.
The exploitation of multiple processors or cores in a shared-memory computer, requires setting up threading mechanisms to assign the workload. In our case, this is rather straightforward as the elements of the domain can be computed separately. However, a linear speedup wasn't achieved, since threads need to be synchronised at the end of each main loop iteration. Indeed, speedup doesn't scale linearly with the number of processors.
Using eight processors, we get a speedup of 4 over the reference implementation and of 2.5 over the mono-core SSE implementation.

{\bf CUDA}.
GPUs (Graphic Processing Unit) are massive multi-core processors (more than 1500 cores in the latest cards) integrated into a single chip. CUDA (Compute Unified Device Architecture) is a practical architecture for general-purpose computing on Nvidia GPUs. Porting our algorithm to CUDA, it requires to specify how to split the work over the cores. Frequent synchronisations are limiting, due to the very large number of cores. We also have to manage the data movements between the main and the GPU memories.
The CUDA implementation is currently in progress, and thereby not yet fully optimised. With a Nvidia GTX680 card with 1536 cores, the preliminary implementation achieves a speedup of 5.6 over the reference implementation.
However, there is still room for considerable optimisation.

\section{Conclusions}

We have developed some theoretical and computing tools for studying bodies motion under self-force, for a specific case. Generalisation to other non-adiabatic orbits are under consideration. Details are given in published and upcoming references.   

\acknowledgments 

V. Allombert, A. Blanchard, A. Carteron, J. Legaux, and S. Limet (LIFO) are acknowledged for their contribution to parallel computing. 

\bibliographystyle{asp2010}
\bibliography{biblio}

 \end{document}